\documentclass[twoside]{article}
\usepackage{Proc_NTSE_16}
\usepackage{bm}       
\usepackage{amsmath}
\usepackage{graphicx,color}

\begin{document}

\newcommand {\nc} {\newcommand}
\nc {\IR} [1]{\textcolor{red}{#1}}
\nc {\IB} [1]{\textcolor{blue}{#1}}
\nc {\IM} [1]{\textcolor{magenta}{#1}}

\thispagestyle{plain}
\publref{elster-ntse16}

\begin{center}
{\Large \bf \strut
Energy Dependent Separable Optical Potentials for (d,p) Reactions
\strut}\\
\vspace{10mm}
{\large \bf 
L. Hlophe$^{a,b}$, Ch. Elster$^{a}$}
\end{center}

\noindent{
\small $^a$\it Institute of Nuclear and Particle Physics,  and
  Department of Physics and Astronomy, \\  Ohio University, Athens, OH 45701, USA} \\
  {\small $^b$\it National Superconducting Cyclotron Laboratory and Department of Physics
  and Astronomy, Michigan State University, East Lansing, MI 48824, USA }

\markboth{
L. Hlophe and Ch. Elster}
{
Energy Dependent Separable Potentials} 

\begin{abstract}
An important ingredient for applications of nuclear physics to e.g. astrophysics or
nuclear energy are the cross sections for reactions of neutrons with rare isotopes. Since
  direct measurements are often not possible, indirect methods like
  $(d,p)$ reactions must be used instead. Those
  $(d,p)$ reactions may be viewed as  effective three-body reactions and
  described with Faddeev techniques. An additional challenge posed by $(d,p)$ reactions
involving
  heavier nuclei is the treatment of the Coulomb force. To avoid numerical complications in
  dealing with the screening of the Coulomb force, recently a new approach using the Coulomb
distorted basis in momentum space was suggested. In order to implement this suggestion
separable representations of neutron- and proton-nucleus optical
  potentials, which are not only complex but also energy dependent, need to be introduced. 
Including excitations of the nucleus in the calculation requires a multichannel
  optical potential, and thus separable representations thereof.
\\[\baselineskip] 
\end{abstract}

\section{Introduction}
Nuclear reactions are an important probe to learn about the structure of unstable nuclei.
  Due to the short lifetimes
  involved, direct measurements are usually not possible. Therefore indirect measurements using
  ($d,p$) reactions have been proposed (see e.g.
  Refs.~\cite{RevModPhys.84.353,jolie,Kozub:2012ka}).
  Deuteron induced reactions are particularly attractive from an experimental perspective,
  since deuterated targets are readily available. From a theoretical perspective they are
  equally attractive because the scattering problem can be reduced to an effective three-body
  problem~\cite{Nunes:2011cv}. Traditionally deuteron-induced single-neutron transfer
  ($d,p$) reactions have been used to study the shell structure in stable nuclei, nowadays
  experimental techniques are available to apply the same approaches to exotic beams (see e.g.
  \cite{Schmitt:2012bt}).
  Deuteron induced $(d,p)$ or $(d,n)$ reactions in inverse kinematics are
  also useful to extract neutron or proton capture rates on unstable nuclei of astrophysical
  relevance. Given the many ongoing experimental programs  worldwide using these reactions, a
  reliable reaction theory for $(d,p)$ reactions is critical.
  
One of the most challenging aspects of solving the three-body problem for nuclear reactions is
  the repulsive Coulomb interaction.
While for very light
  nuclei, exact calculations of (d,p) reactions based on momentum-space Faddeev equations in
  the Alt-Grassberger-Sandhas (AGS)~\cite{ags} formulation can be carried
  out~\cite{Deltuva:2009fp} by using a screening and renormalization
  procedure~~\cite{Deltuva:2005wx,Deltuva:2005cc}, this technique leads to increasing technical
  difficulties when moving to computing (d,p) reactions with
heavier nuclei~\cite{hites-proc}. Therefore, a new formulation of the Faddeev-AGS equations,
  which does not rely on a screening procedure, was presented in
  Ref.~\cite{Mukhamedzhanov:2012qv}. Here the Faddeev-AGS equations are cast in a
  momentum-space  Coulomb-distorted partial-wave representation
  instead of the plane-wave basis.  Thus all operators,  specifically the interactions
  in the two-body subsystems must be evaluated in the Coulomb basis, which is a
  nontrivial task (performed  recently for the neutron-nucleus
  interaction~\cite{upadhyay:2014}).
  The formulation of Ref.~\cite{Mukhamedzhanov:2012qv} requires the interactions in the
  subsystems to be of separable form.

 Separable representations of the forces between constituents forming the subsystems in a
  Faddeev approach have a long tradition, specifically when considering the nucleon-nucleon
  (NN) interaction (see e.g.~\cite{Haidenbauer:1982if,Haidenbauer:1986zza,Entem:2001it}) or
  meson-nucleon interactions~\cite{Ueda:1994ur,Gal:2011yp}. Here the underlying potentials are
Hermitian, and a scheme for deriving separable representations suggested by
  Ernst-Shakin-Thaler~\cite{Ernst:1973zzb} (EST) is well suited, specifically when working in
  momentum space. It has the nice property that the on-shell and half-off-shell transition
matrix elements of the separable representation are exact at predetermined energies, the
  so-called EST support points.
  However, when dealing with neutron-nucleus (nA) or proton-nucleus (pA)
  phenomenological
  optical potentials, which are in general complex to account for absorptive channels that
  are not explicitly treated, as
  well as energy-dependent, extensions of the EST scheme have to be made.

\section{Separable Representation of Single Channel\\ Energy Dependent Optical Potentials}

The pioneering work by Ernst, Shakin and Thaler~\cite{Ernst:1973zzb}
  constructed separable representations of Hermitian potentials. To apply this formalism to
optical potentials, it needs to be extended to handle complex potentials~\cite{Hlophe:2013xca}.
We briefly recall the most
  important features, namely that a separable representation for a complex,
  energy-independent potential
  $U_l$ in a fixed partial wave of orbital angular momentum $l$ is given by~\cite{Hlophe:2013xca}
\begin{equation}
   u_l = \sum\limits_{ij} U_l|\psi_{l,i}^+ \rangle\lambda_{ij}^{(l)}\langle\psi_{l,j}^-|U_l, 
\label{eq:1}
\end{equation}
where $|\psi_{l,i}^+\rangle$ is a solution of the Hamiltonian $H=H_0+U_l$ with outgoing
  boundary conditions at energy $E_i$, and $|\psi_{l,i}^-\rangle$ is a
  solution of the Hamiltonian $H=H_0+U_l^*$ with incoming boundary conditions.
  The energies $E_i$ are referred to as EST support points.
  The free Hamiltonian  $H_0$ has
  eigenstates $|k_i\rangle$ with $k_i^2=2\mu E_i$,
  $\mu$ being  the reduced mass of the neutron-nucleus system.
The EST scheme constrains the matrix $\lambda_{ij}^{(l)}$ with the conditions
\begin{eqnarray}
 \delta_{kj}&=&\sum\limits_{i}\langle\psi_{l,k}^-|U_l|\psi_{l,i}^+\rangle\lambda_{ij}^{(l)} \cr
 \delta_{ik}&=&\sum\limits_{j}\lambda_{ij}^{(l)}\langle\psi_{l,j}^-|U_l|\psi_{l,k}^+\rangle,
\label{eq:2}
\end{eqnarray}
  where the subscript $i=1 \dots N$ indicates the rank of the separable potential.
Those two constraints of Eq.~(\ref{eq:2})  on $\lambda_{ij}^{(l)}$ are an essential 
feature of the EST scheme and
   ensure that at the EST support points $E_i$, both, the original $U$
  and the separable potential $u$, have identical
 wavefunctions or half-shell $t$ matrices. The corresponding separable
  $t$ matrix takes the form           
\begin{equation}
   t_l(E) = \sum\limits_{ij}U_l|\psi_{l,i}^+\rangle\tau^{(l)}_{ij}(E)
  \langle\psi_{l,j}^-|U_l
\label{eq:3}
\end{equation}
  with
\begin{equation}
    \left(\tau^{(l)}_{ij}(E)\right)^{-1}=
\langle\psi_{l,i}^-|U_l-U_lg_0(E)U_l|\psi_{l,j}^+\rangle.
\label{eq:4}
\end{equation}
Here $g_0(E) = (E -H_0 +i \varepsilon)^{-1}$ is the free propagator. The
form factors are given as half-shell $t$-matrices
\begin{equation}
  T_l(E_i)|k_i\rangle\equiv U_l|\psi_{l,i}^+\rangle,
\label{eq:5}
\end{equation}
and are obtained through solving a momentum space Lippmann-Schwinger (LS) equation.                                                                         
However,  when applying the same formulation to an energy-dependent complex potential $U(E)$,
  one obtains
  \begin{equation}
    u_l = \sum\limits_{ij}U_l(E_i)|\psi_{l,i}^+\rangle\lambda^{(l)}_{ij}\langle
  \psi_{l,j}^-|U_l(E_j),
   \label{eq:form4}
  \end{equation}
  with the constraints
  \begin{eqnarray}
  \delta_{kj}&=&\sum\limits_{i}\langle\psi_{l,k}^-|U_l(E_i)|\psi_{l,i}^+\rangle
   \lambda^{(l)}_{ij} \cr
  \delta_{ik}&=&\sum\limits_{j}\lambda^{(l)}_{ij}\langle\psi_{l,j}^-|U_l(E_j)|
  \psi_{l,k}^+\rangle.
\label{eq:form4b}
   \end{eqnarray}
Omitting the partial wave index $l$  the two constraints on $\lambda$
  can be written in matrix form as
\begin{equation}
  \mathcal{U}^t \; \lambda = {\bf 1} =\lambda \;  \mathcal{U},
\label{form5}
\end{equation}
  with
\begin{equation}
  \mathcal{U}_{ij} = \langle\psi_i^-|U(E_i)|\psi_j^+\rangle.
\label{form5b}
\end{equation}
 For a separable potential of rank $N>1$ it is obvious that 
the matrix $\mathcal{U}_{ij}$ is not symmetric
  in the indices $i$ and $j$. This
leads to an asymmetric matrix $\lambda$ and thus a $t$ matrix which violates
  reciprocity.
  Therefore, a different approach must be taken in order to  construct separable
representations for energy-dependent potentials. Here we note that although
  the potential $u$ contains some of the energy dependence of $U(E)$
  through the form factors calculated at the different fixed energy support points $E_i$, 
it has no explicit energy dependence. Thus, this 
 separable construction needs to be considered as  energy-independent
  EST representation.

\begin{figure}[h]
  \centerline{\includegraphics[width=13.0cm]{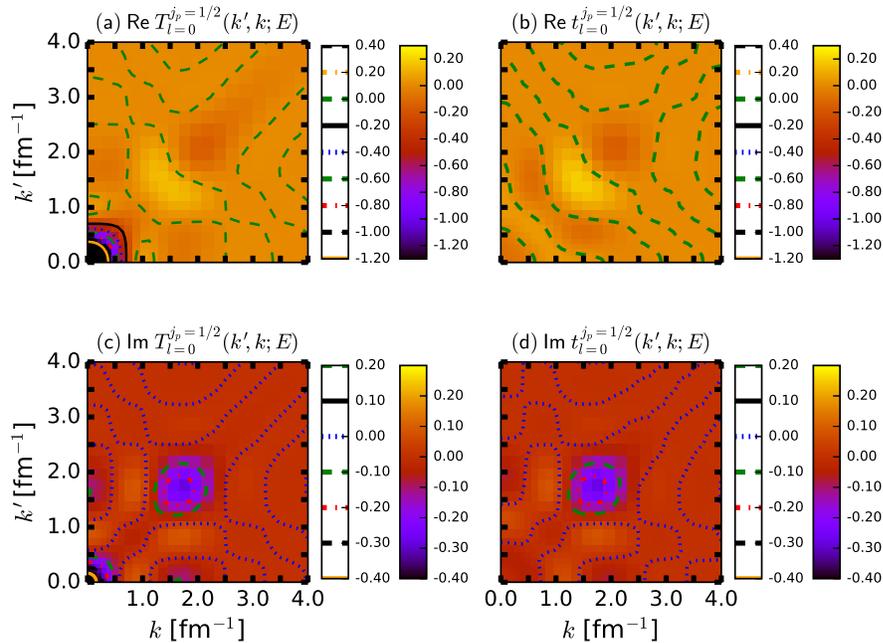}}
  \caption{
The $s$-wave off-shell $t$-matrix elements for the $n$+$^{48}$Ca system calculated
from the CH89 optical potential~\cite{Varner:1991zz} as function of the
off-shell momenta $k$ and $k'$ at 20~MeV
  incident neutron laboratory kinetic energy. Panels (a) and (c) depict the
  real and imaginary $t$-matrix elements corresponding to the CH89 global
  optical potential. The real and imaginary parts of the eEST separable representation of the
off-  shell $t$-matrix are shown in panels (b) and (d). The on-shell momentum is
$k=~0.978$~fm$^{-1}$.
}
\label{fig1}
\end{figure}

A separable expansion for energy-dependent Hermitian potentials was suggested by 
  Pearce~\cite{Pearce:1987zz}. It is straightforward to apply this suggestion to complex
  potentials by using the insights previously gained in~\cite{Hlophe:2013xca}.
  In analogy, we define the EST separable
  representation for complex, energy-dependent potentials (eEST) by allowing
  an explicit energy dependence of the coupling matrix elements $\lambda_{ij}$.
\begin{eqnarray}
 u(E) = \sum\limits_{ij}U(E_i)|\psi_i^+\rangle\lambda_{ij}(E)\langle\psi_j^-|U(E_j),
\label{eq:form9}
\end{eqnarray}
  where the partial wave index $l$ has been omitted for simplicity. In order
  to obtain a constraint on the matrix $\lambda(E)$, we require that the matrix
  elements of the potential $U(E)$ and its separable form $u(E)$ between the
  states $|\psi_i^+\rangle$ be the same at
all energies $E$. This condition ensures that the potentials
  $U(E)$ and $u(E)$ lead to identical wavefunctions at the EST support points,
  just like in the energy-independent EST scheme.

The constraints on $\lambda_{ij}(E)$ become
\begin{eqnarray}
\langle\psi_m^-|U(E)|\psi_n^+\rangle
   &=&\langle\psi_m^-|u(E)|\psi_n^+\rangle\cr
   &=&\sum\limits_{i}\langle\psi_m^-|U(E_i)|\psi_i^+\rangle\lambda_{ij}(E)
   \langle\psi_j^-|U(E_j)|\psi_n^+\rangle.
\label{eq:form10}
\end{eqnarray}
   The corresponding separable $t$-matrix then takes the form
\begin{eqnarray}
   t(E) = \sum\limits_{ij}U(E_i)|\psi_i^+\rangle\tau_{ij}(E)\langle\psi_j^-|U(E_j).
\label{eq:form11}
\end{eqnarray}
  Substituting Eqs.~(\ref{eq:form9})$-$(\ref{eq:form11}) into the LS equation leads to
  constraint for the matrix $\tau(E)$ such that
\begin{equation}
  R(E)\cdot\tau(E) \equiv \mathcal{M}(E),
\label{eq:form12}
\end{equation}
where
\begin{equation}
   R_{ij}(E) = \langle\psi_i^-|U(E_i)|\psi_j^+\rangle-\sum\limits_{n}\mathcal{M}_{in}(E)
  \langle\psi_n^-|U(E_n)\; g_0(E)\; U(E_j)|\psi_j^+\rangle,
\label{eq:form12b}
\end{equation}
  with
\begin{eqnarray}
  \mathcal{M}_{in}(E) \equiv[\mathcal{U}^e(E)\cdot \mathcal{U}^{-1}]_{in}. 
\label{eq:form12c}
\end{eqnarray}
The matrix elements of $\mathcal{U}$ are defined in Eq.~(\ref{form5b}), and
 \begin{equation}
  \mathcal{U}^e_{ij}(E) \equiv \langle\psi_i^-|U(E)|\psi_j^+\rangle.
\label{eq:form13a}
\end{equation}
For  energy-independent potentials $\mathcal{U}^e(E)$ becomes $\mathcal{U}$ and 
the matrix $\mathcal{M}(E)$ is the unit matrix.
The matrix element $\mathcal{U}^e_{ij}(E)$ is explicitly given as
\begin{eqnarray}
 \mathcal{U}^e_{ij}(E) 
 &=&U(k_i,k_j,E)+\int\limits_0^\infty dp p^2 \;T(p,k_i;E_i)\;g_0(E_i,p)\;U(p,k_j,E) \cr
      &+&\int\limits_0^\infty dp p^2\; U(k_i,p,E)\;g_0(E_j,p)\;T(p,k_j;E_j)\\
      &+&\int\limits_0^\infty dp p^2 \int\limits_0^\infty dp' p'^2\; 
      T(p,k_i;E_i)\;g_0(E_i,p)\; U(p,p',E) \;g_0(E_j,p')\;T(p',k_j;E_j)\nonumber,
\label{eq:esep8d0}
\end{eqnarray}
  where $g_0(E,p)=[E-p^2/2\mu+i\varepsilon]^{-1}$.
For the evaluation of $\mathcal{U}^e_{ij}(E)$ for all energies $E$ within the relevant
  energy regime, the form factors $T(p',k_j;E_j)$ are needed at the specified EST
  support points and the matrix elements of the potential $U(p',p,E)$ at
  all energies.  The
explicit derivation of the above expressions is given in
Refs.\cite{Hlophe:2015rqn,LindaHlophe2016}, together with suggestions to simplify the
calculation of $U(p',p,E)$.
\begin{figure}[h]
  \centerline{\includegraphics[width=9.0cm]{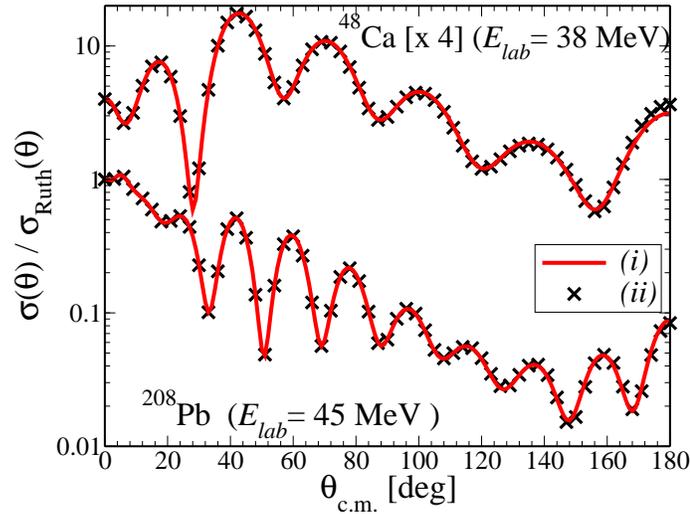}}
  \caption{The unpolarized differential cross section for elastic scattering of
  protons from $^{48}$Ca (upper) and $^{208}$Pb (lower) as function of the
c.m. angle. For $^{48}$Ca the cross section is calculated at a
  laboratory kinetic energy of 38~MeV and is scaled by a factor 4. The calculation for
  $^{208}$Pb is carried out at $E_{lab}$~=~45~MeV. The solid lines ($i$) depict the cross
  section calculated in momentum space based on the rank-5 separable representation of the
  CH89~\cite{Varner:1991zz} phenomenological optical potential, while the crosses
  ($ii$) represent the corresponding coordinate space calculations~\cite{Nunes-private}.
  }
\label{fig2}
\end{figure}
To apply the formulation to proton-nucleus scattering one first realizes that the proton-nucleus
potential consists of the point Coulomb force, $V^c$, together with
  a short-ranged nuclear as well as a short-ranged  Coulomb interaction representing the charge
  distribution of the nucleus, which we refer to as $U^s(E)$. While the point Coulomb
  potential has a simple analytical form, an optical potential is employed
  to model the short-range nuclear potential. The extension of the energy-independent EST
  separable representation to proton-nucleus optical potentials was carried out
  in Ref.~\cite{Hlophe:2014soa}. In that work it was shown that the
  form factors of the separable representation are solutions of the LS equation in the Coulomb
basis, and that  they are obtained using methods introduced in
  Refs.~\cite{Elster:1993dv,Chinn:1991jb}. It was also demonstrated that the extension of the
energy-independent
  EST separable representation scheme to proton-nucleus scattering involves two steps. First,
the nuclear wavefunctions $|\psi_{l,i}^{(+)}\rangle$ are replaced by Coulomb-distorted nuclear
wavefunctions
  $|\psi_{l,i}^{sc~(+)}\rangle$. Second, the free resolvent $g_0(E)$ is replaced by the Coulomb
Green's
  function, $g_c(E) =(E-H_0 -V^c +i\varepsilon)^{-1}$, and third, the energy-dependent scheme 
  must be generalized. 

Upon suppressing the index $l$ we obtain a constraint similar to Eq.~(\ref{eq:form12}),
\begin{eqnarray}
 R^{c}(E)\cdot\tau^{c}(E)
  =\mathcal{M}^{c}_{ij}(E),
\label{pform2}
\end{eqnarray}
with the matrix elements of $R^{c}(E)$ satisfying
\begin{eqnarray}
 R_{ij}^{c}(E) &= &
 \langle\psi_{i}^{sc~(-)}|U^s(E_i)|\psi_{j}^{sc~(+)}\rangle \cr 
  &-& \sum_i\mathcal{M}^{c}_{in}(E)\langle\psi_{n}^{sc~(-)}|U^s(E_n)g_c(E)
U^s(E_j)|\psi_{j}^{sc~(+)}  \rangle.
\label{eq:pform3}
\end{eqnarray}
The matrix $\mathcal{M}^{c}(E)$ is the Coulomb distorted counterpart
  of $\mathcal{M}(E)$ of Eq.~(\ref{eq:form12c}), and is defined as
\begin{equation}
\mathcal{M}^c_{in}(E) = \left[\mathcal{U}^{e,sc}(E)\cdot
({\mathcal{U}^{sc}})^{-1}\right]_{in},
\label{eq:pform4}
\end{equation}
with
\begin{eqnarray} 
  \mathcal{U}^{sc}_{ij}&\equiv& \langle\psi_{i}^{sc~(-)}|U^s(E_i)|\psi_{j}^{sc~(+)}\rangle,\cr
  \mathcal{U}^{e,sc}_{ij}(E)&\equiv&\langle \psi_{k_i}^{sc~(-)} | U^s(E) | \psi_{k_j}^{sc~(+)}
\rangle.
\label{eq:pform4b}
\end{eqnarray}
 If the potential is energy-independent the matrix $\mathcal{M}^c(E)$ becomes a unit matrix just
like $\mathcal{M}(E)$. Further details for the explicit evaluation are given in 
Refs.~\cite{Hlophe:2015rqn,LindaHlophe2016}.

In order to illustrate the quality of the separable representation of energy-dependent optical
potentials for neutron as well as proton elastic scattering, the differential cross sections
for proton scattering off $^{48}$Ca at laboratory kinetic energy 38~MeV and $^{208}$Pb at 45~MeV
are shown in Fig.~\ref{fig2} and compared to the equivalent coordinate space calculations. We
observe that the separable representation provides an excellent description on both cases. The
power of a separable representation based on the EST scheme lies in the choice of the basis, 
namely here the half-shell t-matrices calculated at specific energies. This basis contains a lot
of information about the system considered, and thus only a small number of basis states,
represented by the rank of the separable potential, are needed to have this excellent
representation. 

\begin{figure}[h]
\centerline{\includegraphics[width=9.0cm]{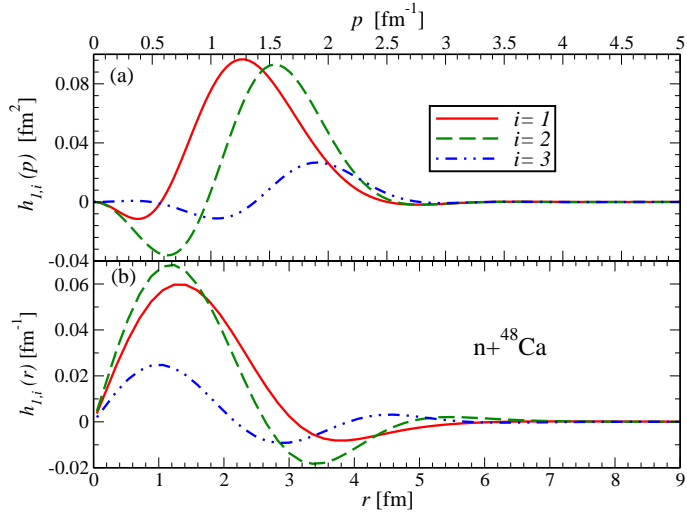}}
\caption{ 
The  $p_{3/2}$ form factors $h_{0,i}$ for the $n+^{48}$Ca system obtained from the  CH89
optical potential~\cite{Varner:1991zz}. Panel (a) illustrates the form factors as function
  of momentum $p$ while panel (b) depicts its Fourier transform as function of the position
coordinate $r$. The indices $i=$~1, 2, and 3 correspond to the support points 5, 21, and
47~MeV.
\label{fig3}
}
\end{figure}

\begin{figure}[ht]
\centerline{\includegraphics[width=9.0cm]{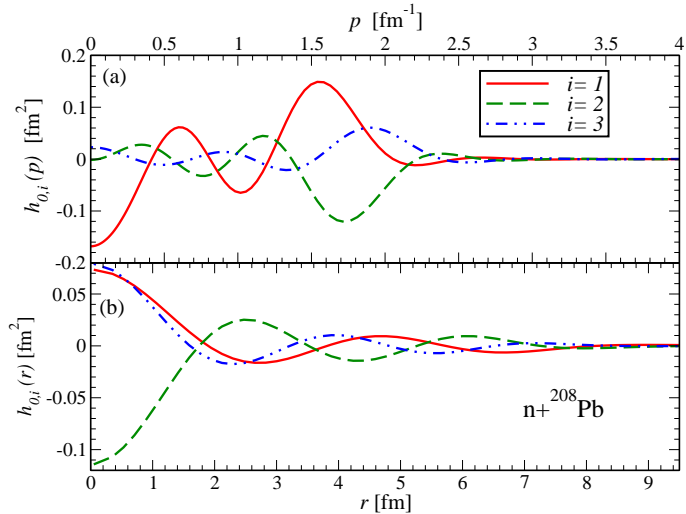}}
\caption{
The $s$-wave form factors $h_{0,i}$ for the $n+^{48}$Pb system obtained from the  CH89 optical
potential~\cite{Varner:1991zz}.  Panel (a) illustrates the form factors as function
 of momentum $p$ while panel (b) depicts its Fourier transform as function of the position
coordinate $r$. The indices $i=$~1, 2, and 3 correspond to the support points 5, 21, and
47~MeV.
}
\label{fig4}
\end{figure}

\section{Coordinate Space Separable Representation of \\ Single Channel Optical Potentials}

\begin{figure}[h]
\centerline{\includegraphics[width=13cm]{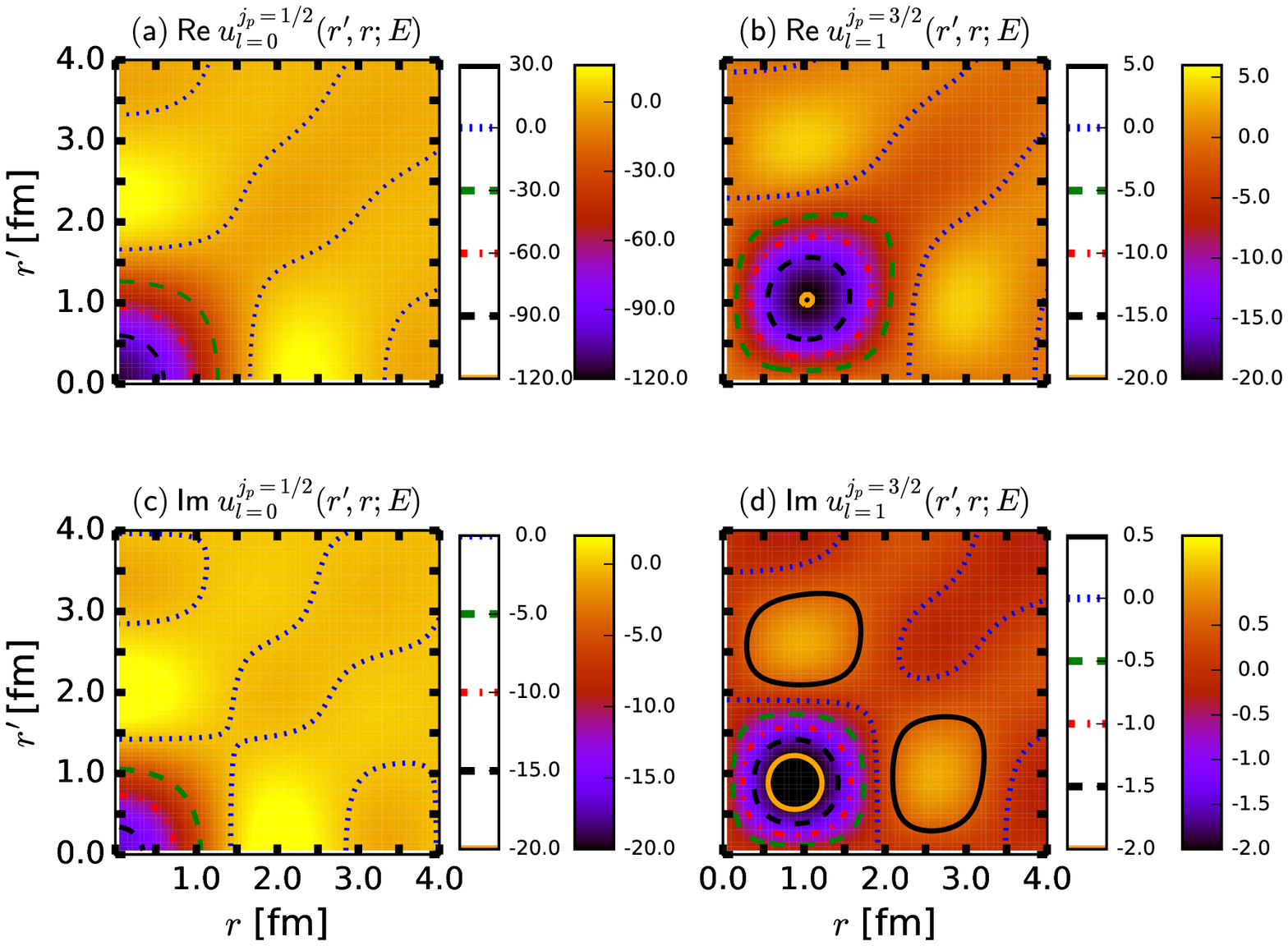}}
\caption{ 
The off-shell potential elements $u_l^{j_p}(r',r,E)$ of the separable representation  
of the CH89 optical potential~\cite{Varner:1991zz} for the $n$+$^{48}$Ca system
as function of the coordinates $r$ and $r'$ at $E=$~20~MeV
  incident neutron laboratory kinetic energy. Panels (a) and (c) depict the
  real and imaginary potential matrix elements for the $s_{1/2}$ partial wave. 
The real and
 imaginary parts of the $p_{3/2}$ separable potential are shown in panels (b) and (d). 
\label{fig5}
}
\end{figure}

The formal scheme for deriving separable representations to Hermitian potentials was given by
Ernst, Shakin, and Thaler in Ref.~\cite{Ernst:1973zzb}, and the application of the of the scheme
to a two-body coordinate space potential representing an s-wave bound and scattering state in
Ref.~\cite{Ernst:1974up}. The authors chose to carry out their construction of the separable
representation in coordinate space, which makes the procedure more cumbersome compared to the
momentum space construction we employ, leading to a momentum space separable representation of either the transition matrix or the potential. 

Since coordinate space techniques have long tradition in nuclear physics, it can be useful to 
consider an EST based separable representation of potentials in coordinate space. 
Separable potentials are
inherently nonlocal. Using the EST formulation leads to a well defined behavior of this
non-locality. However, instead of implementing the EST construction in coordinate space, 
one can carry out the entire scheme in momentum space and then Fourier transform the 
momentum space
result to coordinate space. This is quite simple, since it involves only a one-dimensional
Fourier transform of the form factors. 

To illustrate a coordinate space realization of an EST separable representation, we show
in Fig.~\ref{fig3} the form factors $h_{l,i}$ as function of the momentum $p$ for the
$n+^{48}$Ca system in panel (a) together with their Fourier transformed counterparts in
coordinate space in panel (b). The index $i$ refers to the EST support points used. 
The form factors
are well behaved functions in momentum space as well as coordinate space. In Fig.~\ref{fig4}
the s-wave form factors for the $n+^{208}$Pb system are shown, and we note that for the heavier
nucleus $^{208}$Pb they extend to larger values of $r$ as should be expected considering the
larger size of the heavier nucleus. 

The separable representation of the coordinate space potential in a given partial wave
is obtained by summing over the rank of the potential according to Eq.~(\ref{eq:1}). 
The resulting nonlocal separable coordinate space representation of the CH89 optical potential
 is shown in Fig.~\ref{fig5} for the $n+^{48}$Ca system for the $s_{1/2}$ and $p_{3/2}$ channels.
The non-locality is symmetric in  $r$ and $r'$ as required by reciprocity and its
extension in $r$ and $r'$ is given by the fall-off behavior of the form factors. It also shows a
more intricate behavior than the often employed Perey-Buck Gaussian-type~\cite{Perey:1962}
 non-locality construct. Employing the nonlocal separable representation in solving the
integro-differential Schr\"odinger equation~\cite{Titus:2016gvp} reveals that resulting
coordinate space wavefunction exactly agree with the wavefunctions obtained from solving 
the Schr\"odinger equation with the local CH89 optical potential~\cite{Ross-private}.

\section{Separable Representation of Multi-Channel\\ Energy Dependent Optical Potentials}

To generalize the energy-dependent EST (eEST) scheme to multichannel potentials, 
we proceed analogously to Ref.~\cite{Pieper74} and replace the single-channel
  scattering wavefunctions with their multichannel counterparts, leading to a
  multichannel separable potential
\begin{equation}
  u(E)=\sum\limits_{\rho\sigma}\sum\limits_{ij} 
  \left(\sum\limits_{\gamma JM} U(E_i)\big|\gamma JM\; \Psi_{\gamma\rho,i}^{J(+)}\big\rangle\right) \;
    \lambda_{ij} ^{\rho\sigma}(E)\;
   \left(\sum\limits_{\gamma JM}\big\langle \Psi_{\gamma\sigma,j}^{J(-)}\;\gamma JM\big|U(E_j)\right).
\label{eq:mch1a-1}
\end{equation}
The indices $i$ and $j$ stand for the EST support points.  Using
the definition of a multichannel half-shell $t$ matrix~\cite{Gloecklebook},
\begin{equation}
 T(E_i)|\rho JM \; k_i^\rho  \rangle=\sum_{\gamma} U(E_i)|\gamma\;JM\Psi_{{\gamma\rho}}^{J(+)}\rangle,
\label{eq:nad3c}
\end{equation}
Eq.~(\ref{eq:mch1a-1}) can be recast as 
\begin{equation}
  u(E)=\sum\limits_{JM}\sum\limits_{J'M'}\sum\limits_{\rho\sigma}\sum\limits_{ij} 
   T(E_i)\big|\rho JM\; k_i^\rho\big\rangle \;\lambda_{ij}^{\rho\sigma}(E)\;
   \big\langle k_j^\sigma\;\sigma J'M'\big|T(E_j).
\label{eq:mch21a}
\end{equation}
To determine the constraint on $u(E)$,
   we first generalize
   the matrices ${\cal U}^{e}(E)$ and ${\cal U}$ to
multichannel potentials. This is accomplished by replacing
   the single-channel scattering states by the multichannel
   scattering states so that
\begin{eqnarray}
  {\cal U}^{e,\alpha\beta}_{mn}(E)&\equiv&\sum\limits_{\gamma\nu}\big\langle
\Psi_{\gamma\alpha,m}^{J(-)}\;\gamma JM | U(E)|\nu JM\;\Psi_{\nu\beta,n}^{J(+)}\rangle,\cr
    &=&\sum\limits_{\gamma\nu}\big\langle \Psi_{\gamma\alpha,m}^{J(-)}|
    U_{\gamma\nu}^J(E)|\Psi_{\nu\beta,n}^{J(+)}\rangle,
\label{eq:mch21b0}
\end{eqnarray}
and
\begin{equation}
    {\cal U}^{\alpha\beta}_{mn}\equiv{\cal U}^{e,\alpha\beta}_{mn}(E_m)
     =\sum\limits_{\gamma\nu}\big\langle \Psi_{\gamma\alpha,m}^{J(-)}\big|
    U_{\gamma\nu}^J(E_m)\big|\Psi_{\nu\beta,n}^{J(+)}\big\rangle.
\label{eq:mch21b1}
\end{equation}
The $J$ dependence of matrix elements ${\cal U}^{e,\alpha\beta}_{mn}(E)$
  and ${\cal U}^{\alpha\beta}_{mn}$ is omitted for simplicity.
  One one hand, Eq.~(\ref{eq:mch21b1}) shows that the matrix
  ${\cal U}$ depends only on
  the support energies $E_m$ and $E_n$. On other hand, we see
  from Eq.~(\ref{eq:mch21b0}) that ${\cal U}^{e}(E)$ depends
  on the projectile energy $E$ as well as the support energies.
  The  constraint on the separable potential is obtained
  by substituting the multichannel matrices ${\cal U}^e$ and ${\cal U}$ into
  Eq.~(\ref{eq:form10}) leading to
\begin{eqnarray}
  {\cal U}^{e,\alpha\beta}_{mn}(E)
   &=& \sum\limits_{\rho\sigma}\sum\limits_{ij}\big({\cal U}^t\big)^{\alpha\rho}_{mi}\;
   \lambda_{  ij}^{\rho\sigma}(E)\;{\cal U}^{\sigma\beta}_{jn},\cr
  &=&\left[{\cal U}^t\cdot\lambda(E)\cdot {\cal U} \right]_{mn}^{\alpha\beta}. 
\label{eq:mch21b2}
\end{eqnarray}
To evaluate the separable multichannel $t$ matrix, we
  insert Eqs.~(\ref{eq:mch21a})-(\ref{eq:mch21b2}) into
  a multi-channel LS equation  and obtain
\begin{eqnarray}
t(E)&=&\sum\limits_{\rho\sigma}\sum\limits_{ij} 
    \left(\sum\limits_{\gamma JM} U(E_i)\big|\gamma JM\;
\Psi_{\gamma\rho,i}^{J(+)}\big\rangle\right) \;\tau_{ij} ^{\rho\sigma}(E)\;
   \left(\sum\limits_{\gamma JM}\big\langle \Psi_{\gamma\sigma,j}^{J(-)}\;\gamma
 JM\big|U(E_j)\right) \cr
   &=&\sum\limits_{JM}\sum\limits_{J'M'}\sum\limits_{\rho\sigma}\sum\limits_{ij} 
     T(E_i)\big|\rho JM\; k_i^\rho\big\rangle \;\tau_{ij}
    ^{\rho\sigma}(E)\;
   \big\langle k_j^\sigma\;\sigma J'M'\big|T(E_j).
\label{eq:mch21c}
\end{eqnarray}
The coupling matrix elements $\tau_{ij}^{\rho\sigma}(E)$ fulfill
\begin{eqnarray}
    R(E)\cdot\tau(E)=\mathcal{M}(E),
\label{eq:mch24a}
\end{eqnarray}
where
\begin{eqnarray}
R^{\rho\sigma}_{ij}(E)&=&\Big\langle
k_i^{\rho}\Big|~T_{{\rho\sigma}}^J(E_i)~+~\sum\limits_{  {\beta}}
    T_{{\rho\beta}}^J(E_i)G_{{\beta}}(E_j)T_{ {\beta\sigma}}^J(E_j)\Big| k_j^{\sigma}
    \Big\rangle\cr\cr
   &-&~\sum\limits_{ {\beta\beta}'}\sum\limits_{n}{\cal M}_{in}^{ {\rho\beta}}\langle
k_n^{\beta}\Big|
    T_{ {\beta\beta'}}^J(E_n)G_{ {\beta'}}(E)T_{ {\beta'\sigma}}^J(E_j)
   \Big| k_j^{\sigma}\Big\rangle,
\label{eq:mch24b}
\end{eqnarray}
and
\begin{equation}
 {\cal M}^{\rho\sigma}_{ij}(E)=\left[{\cal U}^e(E)\cdot{\cal U}^{-1}\right]^{\rho\sigma}_{ij}.
\label{eq:mch25c}
\end{equation}
The expression for the matrix $R^{\rho\sigma}_{ij}(E)$ is analogous to
   the one for the single-channel case except for the extra channel indices. 

\begin{figure}[h]
\includegraphics[width=12cm]{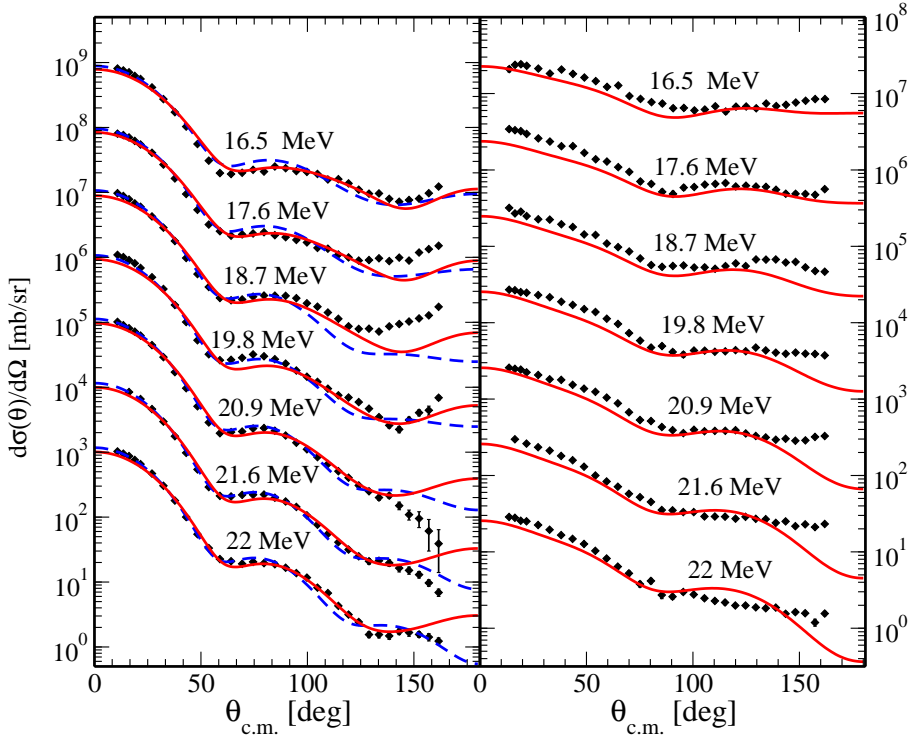}
\caption{ The differential cross sections for scattering in the  $n+^{12}$C system
 computed at different incident neutron energies with the eEST separable representation of the
  Olsson~89 DOMP~\cite{Olsson:1989npa} (solid lines). 
  The left hand panel shows the differential cross section for elastic scattering, while  the right 
  hand panel depicts the differential cross section for inelastic scattering to
  the $2^+$ state of $^{12}$C.  The dashed
  lines indicate cross sections computed with the spherical
  Olsson~89~\cite{Olsson:1989npa} OMP.
  The filled diamonds represent  the data taken from Ref.~\cite{Olsson:1989npa}.
  The cross sections are scaled up by multiples of 10.
  The results at 21.6~MeV are multiplied by 10, those at 20.9~MeV
  are multiplied by 100, etc.
\label{fig:fig6}
}
\end{figure}
To illustrate the implementation of the multichannel eEST separable
  representation scheme, we consider the scattering of neutrons from the
  nucleus $^{12}$C. The $^{12}$C nucleus possesses  selected excited
  states, with the first and second  levels having
   $I^\pi=2^+$  and $I^\pi=4^+$ and being  located at 4.43 and 14.08~MeV above the $0^+$ ground
state. The collective rotational model~\cite{ThompsonNunes} is
   assumed to the coupling between the ground state and these excited states.
   We consider here elastic scattering
   and inelastic scattering to the $2^+$ rotational state.
To test the multichannel eEST separable representation we use the 
deformed optical potential model (DOMP)
 derived by Olsson et {\it al.}~\cite{Olsson:1989npa} and fitted to elastic and inelastic 
scattering data between 16 and 22~MeV laboratory kinetic energy. 
In Fig.~\ref{fig:fig6} the differential
 cross sections for elastic and inelastic scattering for the $n+^{12}$C system are shown
at various incident neutron energies. The left hand panel shows the
       differential cross section for elastic scattering, and  the right hand panel the
   differential cross section
   for inelastic scattering to the $2^+$ state of $^{12}$C. The support points are at
$E_{lab}=$~6 and 40~MeV. The separable representation describes both
  differential cross sections very well. In addition, it is
  in good agreement with the coupled-channel calculations
     shown in Fig.~1 of Ref.~\cite{Olsson:1989npa}.
The dashed lines indicate cross sections computed with the spherical
       Olsson~89~\cite{Olsson:1989npa} OMP.

\section{Summary and Outlook}
In a series of steps we developed the input that will serve as a  basis for Faddeev-AGS
three-body calculations of
$(d,p)$ reactions, which will not rely on the screening of the
Coulomb force. To achieve this, Ref.~\cite{Mukhamedzhanov:2012qv} formulated the
Faddeev-AGS equations in the Coulomb basis using separable interactions in the two-body
subsystems.We developed separable representations of phenomenological optical
  potentials of Woods-Saxon type for neutrons and protons. First we concentrated on
  neutron-nucleus optical potentials and generalized the Ernst-Shakin-Thaler (EST)
  scheme~\cite{Ernst:1973zzb} so that it can be applied to complex and energy-dependent
 optical potentials~\cite{Hlophe:2013xca,Hlophe:2015rqn}. In order to consider proton-nucleus
optical potentials,
  we further extended the EST scheme so that it can be applied to the scattering of charged
  particles with a repulsive Coulomb force~\cite{Hlophe:2014soa}. Finally we extended the
EST formulation to incorporate multi-channel optical potentials~\cite{Hlophe:2016isw}.

The results demonstrate, that separable representations based on a generalized EST scheme
reproduce standard coordinate space
  calculations of neutron and proton scattering cross sections very well. We also showed that
from momentum space separable representations corresponding coordinate space representations can
be obtained using Fourier transforms of the form factors. From those solutions,  observables for
$(d,p)$
  transfer reactions using a Faddeev-AGS formulation  should be readily calculated. 
Work along these lines is in progress.

\vspace{2mm}
  \begin{center}{\bf Acknowledgments} \end{center}
  \vspace{-1mm}
This material is based on work  in part supported
  by the U.~S.  Department of Energy, Office of Science of Nuclear Physics
  under contract
  No. DE-FG02-93ER40756 with Ohio University. The authors thank F.M. Nunes and A. Ross for 
fruitful discussions.

  
\bibliographystyle{h-physrev5}

\bibliography{coulomb}

\end{document}